\documentclass[aps,pra,twocolumn,10pt,amsmath,showpacs]{revtex4-1}
\usepackage[english]{babel}
\usepackage[latin1]{inputenc}
\usepackage{amsfonts}
\usepackage{amsthm}
\usepackage{mathrsfs}
\usepackage[pdftex]{color,graphicx}

\begin{document}
\title{Hybrid and optical implementation of the Deutsch-Jozsa algorithm}
\date{\today}
\author{Luis~A.~Garc\'ia}
\email{anib-gar@uniandes.edu.co}
\author{Jagdish~R.~Luthra}
\email{jluthra@uniandes.edu.co}
\affiliation{Departament of Physics, Universidad de los Andes, Bogot\'a 111711, Colombia}

\begin{abstract}
	A hybrid model of the Deutsch-Jozsa algorithm is presented, inspired by the proposals of hybrid computation by S. Lloyd and P. van Loock \textit{et. al}. The model is based on two observations made about both the discrete and continuous algorithms already available. First, the Fourier transform is a single-step operation in a continuous-variable (CV) setting. Additionally, any implementation of the oracle is nontrivial in both schemes. The steps of the computation are very similar to those in the CV algorithm, with the main difference being the way in which the qunats, or quantum units of analogic information, and the qubits interact in the oracle. Using both discrete and continuous states of light, linear devices, and photo-detection, an optical implementation of the oracle is proposed. For simplicity, infinitely squeezed states are used in the continuous register, whereas the optical qubit is encoded in the dual-rail logic of the KLM protocol. The initial assumption of ideal states as qunats will be dropped to study the effects of finite squeezing in the quality of the computation.
\end{abstract}

\maketitle

\section{Introduction}
The model proposed by Knill, Laflamme and Milburn \cite{KLM01,KLM00} presents an efficient way to implement a quantum computer using optical qubits, single-photon sources, linear devices and photo-detection, known as the KLM protocol. This well established model uses nonlinearities hidden in the measurement process to produce the interactions needed between the optical modes, together with offline state preparation and teleportation to enhance the probability of success of the conditional gates of the protocol, and error correction codes, achieving a scheme that is both fault-tolerant and robust against errors \cite{LOQC05,LOQCrev07}.

On the other hand, the unconditional nature in which entanglement with continuous variables can be obtained, and thus the \emph{unconditionalness} of the operations performed over CV together with the need of just single-mode nonlinearities for universal quantum computation, are two of the most prominent reasons to adopt this setting over a discrete-variable one \cite{CVQCrev05,CVQC99}. Besides, the idea of studying quantum algorithms, generally designed for qubits, in a CV setting is at the same time challenging and of great importance for the development of quantum information theory. This generalization may bring into light new algorithms with a more natural formulation using CVs. For example, any algorithm based on the quantum Fourier transform, such as the Deutsch-Jozsa algorithm or Shor's algorithm, is easily performed in a continuous setting, where such operation is a single-step operation \cite{CVGK02}. 

Focusing on the Deutsch-Jozsa algorithm, it can be seen that its original discrete formulation \cite{DJ92} requires the production of superpositions of several qubit states via a quantum Fourier transform, a task that needs a lot of resources and computational power, even for a small number of qubits \cite{Nielsen}. The translation of the algorithm into a CV setting seems more feasible and practical, due to the ability to produce the superpositions needed with a very low number of resources, nonetheless, how to implement the oracle in such continuous scheme is still an unresolved problem \cite{CVDJ02}.

However, we can take advantage of both schemes by means of some sort of \emph{hybrid} gates, i.e. a set of transformations that acts on both qubits and CV states \cite{Hybr00}. In this paper we use a hybrid model of computation, in an all-optical setting, giving an explicit optical circuit for the action of the oracle on both registers involved in the computation. Such model is based on proposals by S.~Lloyd \cite{Hybr00} and P.~van~Loock \textit{et. al.} \cite{Hybr08}, but differs from them by utilizing only linear optical devices, single-photon sources, offline squeezing and photo-detectors. In the first part of the work, we use idealized infinitely squeezed states as the computational CV basis for simplicity in our calculations. By the end, we shall drop such restriction to discuss the effects of realistic Gaussian states (that is states with finite squeezing) throughout the computational process.

The structure of the paper is as follows. First, in Section \ref{sec:DJfund} we give a quick review of the original algorithm and its implementation in the continuous-variable case. Section \ref{sec:HybrDJ} presents the algorithm in a hybrid and all-optical setting, paying special attention to the explicit implementation of the oracle. Section \ref{sec:gauss} considers imperfect CV states in the hybrid model of the algorithm, in order to analyze the possible distortions at the end of the computation. Finally, Section \ref{sec:concl} closes the paper with some concluding remarks.

\section{Fundamentals of the Deutsch-Jozsa algorithm} \label{sec:DJfund}
In this section we present a review of the basic aspects of the Deutsch-Jozsa problem according to its original formulation. Then, the explicit algorithm is given for CVs as presented in Ref. \cite{CVDJ02}.

	\subsection{Overview of the algorithm} \label{sec:DJov}
	In the Deutsch-Jozsa problem \cite{DJ92}, we are given a black box quantum computer, known as the \emph{oracle}, that implements the function $f:\{0,1\}^n\rightarrow \{0,1\}$. We are \emph{promised} that the function is either \emph{constant} (0 on all inputs or 1 on all inputs) or \emph{balanced} (returns 1 for half of the input domain and 0 for the other half;) the task then is to determine if $f$ is constant or balanced by using the oracle. Although it is of little practical use, it is one of the first examples of a quantum algorithm that is exponentially faster than any possible deterministic classical algorithm. It also provided inspiration for Shor's and Grover's algorithm \cite{Shor94,Grover96}, two of the most revolutionary quantum algorithms.
	
	For a conventional deterministic algorithm where $n$ is the number of bits, $2^{n-1} + 1$ evaluations of $f$ will be required in the worst case. That is to say that in order to prove that $f$ is constant, just over half the set of outputs must be evaluated and found to be identical (remembering that the function is guaranteed to be either balanced or constant, not somewhere in between.) The best case occurs when the function is balanced and the first two output values chosen happen to be different. For a conventional randomized algorithm, a constant $k$ evaluations of the function suffices to produce the correct answer with a high probability (failing with probability $\varepsilon\leq 1/2^{k-1}$.) However, $k=2^{n-1}+1$ evaluations are still required if we want an answer that is always correct. The Deutsch-Jozsa algorithm produces an answer that is \emph{always} correct with a single evaluation of $f$.

	\subsection{The algorithm with CVs} \label{sec:CVDJ}
	Implementing a superposition for a large number of qubits in a discrete setting, although possible, seems non-practical. This represents an important drawback for the implementation of many quantum algorithms that highly depend on such operation. The Deutsch-Jozsa algorithm is one of them and in the present section we show an alternative to apply the algorithm using qunats. In this case, the advantage of using a continuous setting comes from the low computational resources needed to implement a continuous superposition; in fact, this is easily achieved through a Fourier gate on the incident qunat state \cite{CVGK02}.
	
	In the CV case of the Deutsch-Jozsa algorithm \cite{CVDJ02}, Alice and Bob play the following game: Alice chooses a random \emph{real} number $q$ between $-\infty$ and $+\infty$ (or between $-L$ and $+L$ in practice,) and then, she asks Bob to evaluate the function $f(q)$ with possible outcomes 0 and 1. Notice that this game implies that, classically, Alice could only discover the nature of the function with total certainty if she asks Bob an infinite number of times! A quantum algorithm, however, only requires one evaluation of the function to determine if it is either constant or balanced, thus providing a significant speedup over any classical algorithm.
	
	\begin{figure}[htb]
		\centering
		\includegraphics[width=0.45\textwidth]{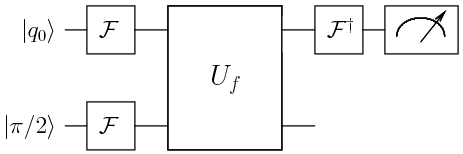}
		\caption{Quantum circuit for the Deutsch-Jozsa algorithm with continuous variables.}
		\label{fig:CVDJ}
	\end{figure}
	
	To store her query, Alice has a qunat register to store her query in, and so does Bob, who stores the answer of the oracle in it. The quantum circuit representing the action of the algorithm on both qunat registers is shown in figure \ref{fig:CVDJ}. Initially, the states are prepared in the ideal CV states $\left|q_0\right\rangle$ and $\left|\pi/2\right\rangle$ for Alice and Bob, respectively, both in the $q$-quadrature basis of the electromagnetic field. Afterwards, they both produce a superposition in each register using Fourier gates, producing the state
	\begin{equation} \label{eq:CVDJ1}
		\frac{1}{\pi}\int\mathrm{d}q\:\mathrm{d}q'e^{2iq_0q+i\pi q'}\left|q\right\rangle\left|q'\right\rangle.
	\end{equation}
	The oracle then performs the nondemolition operation between the qunat states $\left|q_1\right\rangle\left|q_2\right\rangle \rightarrow \left|q_1\right\rangle\left|q_2+f(q_1)\right\rangle$, where the information about $f$ is stored in the second variable.
	{\setlength\arraycolsep{2pt}
	\begin{eqnarray} \label{eq:CVDJ2}
		& & \frac{1}{\pi}\int\mathrm{d}q\:\mathrm{d}q'e^{2iq_0q+i\pi q'}\left|q\right\rangle \left|q'+f(q)\right\rangle \nonumber\\
		& = & \frac{1}{\sqrt{\pi}}\int\mathrm{d}q\:e^{2iq_0q}\left(-1\right)^{f(q)}\left|q\right\rangle \mathcal{F}\left|\pi/2\right\rangle.
	\end{eqnarray}}
	At this point, Bob's register has done its work and remains unchanged until the end of the computation. We apply one last (inverse) Fourier gate to Alice's register before the measurement stage
	\begin{equation} \label{eq:CVDJ3}
		\left|Q\right\rangle=\frac{1}{\pi}\int\mathrm{d}q\:\mathrm{d}q'e^{2iq(q_0-q')}(-1)^{f(q)} \left|q'\right\rangle.
	\end{equation}
	Finally, Alice makes a projective measurement in her qunat state. She projects it onto the initial state $\left|q_0\right\rangle$, using the operator
	\begin{equation} \label{eq:CVDJ4}
		\hat{\Pi}_{\Delta q_0}\equiv\int^{q_0+\Delta q_0/2}_{q_0-\Delta q_0/2}\mathrm{d}k\left|k\right\rangle\left\langle k\right|,
	\end{equation}
	which takes into account certain spread $\Delta q_0$ in the measurement process. The state $\left|Q\right\rangle$ becomes then
	\begin{equation} \label{eq:CVDJ5}
		\hat{\Pi}_{\Delta q_0}\left|Q\right\rangle=\frac{1}{\pi}\int^{q_0+\Delta q_0/2}_{q_0-\Delta q_0/2}\mathrm{d}k \int\mathrm{d}q\:e^{2iq(q_0-k)}(-1)^{f(q)}\left|k\right\rangle.
	\end{equation}
	Let us first suppose the function is constant. In that case, $f(q)=\pm1$ necessarily, so the projection in (\ref{eq:CVDJ5}) can be simplified to
	\begin{equation} \label{eq:CVDJ6}
		\hat{\Pi}_{\Delta q_0}\left|Q\right\rangle=\pm\left|q_0\right\rangle\mathcal,
	\end{equation}
	and then Alice measures with total certainty her initial quadrature state $\left|q_0\right\rangle$. In the other case, when the function is balanced, the output quadrature would have any value but $q_0$, yielding a zero measurement.
	
	In practice, the preparation of nonideal computational states and finite precision in measurement pose severe problems for the experimental realization of this scheme, specially because the constant or balanced nature of the function would not be as straightforward to differentiate as in the idealized case presented above, the outcome of the measurement would not be as certain. Additionally, this model does not address the question of the implementation of the oracle, but suggests that using some kind of \emph{hybrid} setting (i.e. one that combines discrete and continuous states,) one may improve the algorithm.

\section{Deutsch-Jozsa algorithm in a hybrid setting} \label{sec:HybrDJ}
In the hybrid computer proposed in Ref. \cite{Hybr00}, necessary and sufficient conditions are given to build up a universal set of gates operating in both a qubit and a CV state. Actually, the construction of such set is completed with just one interaction gate between both states, because universality is achieved by the repeated application of such interaction gate and the single-mode gates of each setting.

\begin{figure}[htb]
	\centering
	\includegraphics[width=0.45\textwidth]{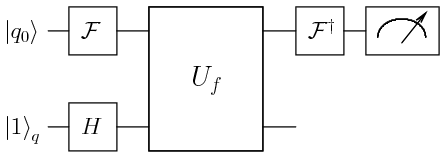}
	\caption{Quantum circuit for the Deutsch-Jozsa algorithm in a hybrid setting.}
	\label{fig:HybrDJ}
\end{figure}

Let us consider the Deutsch-Jozsa problem in a hybrid setting. We use a qunat state to store Alice's query and a qubit to store the result of Bob's evaluation. The quantum circuit is given in figure \ref{fig:HybrDJ}. We keep the continuous part to implement the Fourier gate, which is much more efficient than the discrete superposition achieved with consecutive Hadamard gates \cite{Nielsen}; but we also use a discrete register in which it is easier to store the answer given by the oracle.

	\subsection{Optical implementation of the algorithm} \label{sec:OptDJ}
	The all-optical setting we present here starts with an ideal\footnote{As said before, this consideration simplifies our initial analysis, but it will be dropped in the next section.} qunat state $|q_0\rangle$ in Alice's register, and a dual-rail qubit in Bob's register. The latter requires two paths or \emph{rails} through which a single photon travels, a $|0\rangle_q$ qubit is encoded if the lower path is populated $|0,1\rangle$, and a $|1\rangle_q$ qubit is obtained when the upper rail carries the photon \cite{LOQC05}. For the particular case of this algorithm, Bob produces a $|1\rangle_q$ qubit before the computation.
	
	\begin{figure}[htb]
		\centering
		\includegraphics[width=0.45\textwidth]{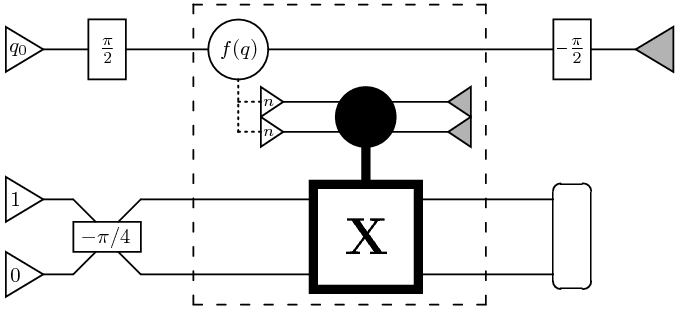}
		\caption{Optical circuit of the hybrid Deutsch-Jozsa algorithm. The explicit implementation of the oracle is shown.}
 		\label{fig:HybrOra}
	\end{figure}
	
	Having prepared both registers, Alice and Bob produce superpositions on each of them. For the CV state, she uses a phase shifter with $\varphi=\pi/2$ to implement the Fourier gate \cite{CVQC99,CVGK02}. On the other hand, Bob performs the Hadamard gate using a 50--50 beam splitter ($\theta=\pi/4$.) The optical circuit is shown explicitly in figure \ref{fig:HybrOra}. The resulting state just before the oracle operation is then
	\begin{equation} \label{eq:HybrDJ1}
		\left(\frac{1}{\sqrt{\pi}}\int\mathrm{d}q\:e^{2iq_0q}\left|q\right\rangle\right)\otimes \frac{1}{\sqrt{2}}(\left|0\right\rangle_q-\left|1\right\rangle_q).
	\end{equation}
	
	In this setting, we require that the action of the oracle transforms the state (\ref{eq:HybrDJ1}) into
	\begin{equation} \label{eq:HybrDJ2}
		\left(\frac{1}{\sqrt{\pi}}\int\mathrm{d}q\:e^{2iq_0q}(-1)^{f(q)}\left|q\right\rangle\right)\otimes \mathbf{H}\left|1\right\rangle_q,
	\end{equation}
	in analogy to the action of the oracle in the CV setting. For now, it is all we are saying about the oracle, the complete analysis is given in the next section.
	
	As usual, Bob's register is left unchanged after the oracle operation, and Alice performs one last superposition using an inverse Fourier gate (via a phase shifter with $\varphi=-\pi/2$.) Actually, these final steps are identical to those in the CV implementation, i.e. steps (\ref{eq:CVDJ3}) through (\ref{eq:CVDJ6}). Once more, if the function is constant, Alice measures her initial state $|q_0\rangle$ with total certainty. In contrast, she measures zero if the function is balanced.

	\subsection{Operation of the Oracle} \label{sec:Oracle}
	Now, the crucial step of the algorithm occurs in the oracle, where the registers interact with each other. The hybrid nondemolition operation required in this point is given by
	\begin{equation} \label{eq:orac}
		\left|q\right\rangle\left|x\right\rangle_q\longrightarrow\left|q\right\rangle\left|x\oplus f(q)\right\rangle_q,
	\end{equation}
	where $|q\rangle$ is an infinitely squeezed state, and $|x\rangle_q$ is an optical qubit. The operation $x\oplus f(q)$ means addition modulo 2. Now we make an important observation. As $f(q)$ can only be 0 or 1, this operation looks very much like a \textsc{cnot} gate in the discrete setting. Let us use this intuition to construct the oracle with linear optical devices, photon sources and photo-detection, as shown in the boxed section of figure \ref{fig:HybrOra}.
	
	After the superposition stage, the oracle evaluates the function once in the superposition state created by Alice. The result (0 or 1) is used to create another dual-rail \emph{control} qubit, which is conditioned by the feed-forward given by the oracle:
		\begin{enumerate}
			\item If $f(q)=0$, we create a state $\left|0\right\rangle_c=\left|0,1\right\rangle$, that is, the photon source of the top rail does not produce anything, and the source of the bottom path creates a single photon,
			
			\item or if $f(q)=1$, we produce a single photon in the upper arm and no photons in the other, obtaining the qubit $\left|1\right\rangle_c=\left|1,0\right\rangle$.
		\end{enumerate}
		
		This method allows the qunat to pass the oracle unchanged, the only operation made on it transmitted classically to the ancillary control qubit. Therefore, the operation (\ref{eq:orac}) becomes simply a \textsc{cnot} gate between qubits
		\begin{equation} \label{eq:orac2}
			\left|x\right\rangle_c\left|y\right\rangle_q \longrightarrow \left|x\right\rangle_c\left|y\oplus x\right\rangle_q,
		\end{equation}
		with the information about $f$ hidden in the value of $x$. The optical implementation of a \textsc{cnot} gate can be efficiently performed using the KLM protocol of computation together with error correction codes \cite{KLM01,KLM00}.
		
		Remember that Bob's register is in the state $(\left|0\right\rangle_q-\left|1\right\rangle_q)/\sqrt{2}$ before the oracle action. The \textsc{cnot} gives us then the desired result: if $f(q)=0$, Bob's qubit is left unchanged, whereas if $f(q)=1$, it gains a sign. The net action is then a phase $(-1)^{f(q)}$, as expected.
	
\section{Effects of Gaussian states in the computation} \label{sec:gauss}
Instead of using the unphysical, infinitely squeezed states, we replace them for the more realistic Gaussian states, that represent Gaussian fields such as the vacuum state $|0\rangle$ or states with finite squeezing $\mathcal{S}(s)|0\rangle$. The latter are produced with nonlinear optical processes such as optical parametric amplification, which allows us to generate squeezed vacuum states \cite{CVclus09}. The wave function of a general Gaussian state is given by the distribution of variance $s^2/2$
\begin{equation} \label{eq:gaussW}
	G_s(q)=\left\langle q\right.\!\left|s\right\rangle=(\pi s)^{-1/2}e^{-q^2/s^2},
\end{equation}
and thus, the vacuum squeezed state $|s\rangle=\mathcal{S}(s)|0\rangle$ can be written as
\begin{equation} \label{eq:gaussS}
	\left|s\right\rangle=\int\mathrm{d}q\:G_s(q)\left|q\right\rangle.
\end{equation}
The parameter $s$ is related to the squeezing parameter through $r=\ln(s)$, and thus we shall call it also the squeezing parameter.

Now we can follow the steps of the algorithm with these realistic states in Alice's CV register. The initial state is then $|s\rangle\otimes|1\rangle_q$ or
\begin{equation} \label{eq:GaussDJ1}
	\left((\pi s)^{-1/2}\!\int\mathrm{d}q\:e^{-q^2/s^2}\left|q\right\rangle\right)\otimes\left|1\right\rangle_q.
\end{equation}
Both the continuous and discrete superpositions are performed in the respective register, transforming the state into
\begin{equation} \label{eq:GaussDJ2}
	\left(\frac{1}{\sqrt{\pi}}\int\mathrm{d}q\:e^{-s^2q^2}\left|q\right\rangle\right)\otimes \frac{1}{\sqrt{2}}(\left|0\right\rangle_q-\left|1\right\rangle_q).
\end{equation}
The oracle operates and, as seen in the previous section, the resulting state gains a phase $(-1)^{f(q)}$
\begin{equation} \label{eq:GaussDJ3}
	\left(\frac{1}{\sqrt{\pi}}\int\mathrm{d}q\:e^{-s^2q^2}(-1)^{f(q)}\left|q\right\rangle\right)\otimes \mathbf{H}\left|1\right\rangle_q.
\end{equation}
Finally, Alice implements the final Fourier gate in her register (Bob's register is left unchanged and thus we drop it), producing the state
\begin{equation} \label{eq:GaussDJ4}
	\left|\mathbf{S}\right\rangle=\frac{1}{\pi}\int\mathrm{d}q\:\mathrm{d}q'\:e^{-s^2q^2-2iqq'}(-1)^{f(q)}\left|q'\right\rangle,
\end{equation}
right before the measurement stage. Alice wants to measure her initial state, so she makes a projective measurement with a little spread $\Delta s$ around $s$
\begin{equation} \label{eq:GaussDJ5}
	\hat{\Pi}_{\Delta s}\equiv\int^{s+\Delta s/2}_{s-\Delta s/2}\mathrm{d}t\left|t\right\rangle\left\langle t\right|,
\end{equation}
obtaining the state
\begin{equation} \label{eq:GaussDJ6}
	\hat{\Pi}_{\Delta s}\left|\mathbf{S}\right\rangle=\frac{1}{\pi}\int^{s+\Delta s/2}_{s-\Delta s/2}\mathrm{d}t \int\mathrm{d}q(-1)^{f(q)}e^{-(s^2+t^2)q^2}\left|t\right\rangle.
\end{equation}
Again, if the function is constant, the term $(-1)^{f(q)}$ becomes $\pm1$, leaving another Gaussian integral to solve, yielding
\begin{equation} \label{eq:GaussDJ7}
	{}=\pm\frac{1}{\pi}\int\mathrm{d}q\int^{s+\Delta s/2}_{s-\Delta s/2} \frac{\mathrm{d}t}{t^2}\left(1+\frac{s^2}{t^2}\right)^{-1/2}e^{-q^2/t^2}\left|q\right\rangle,
\end{equation}
after writing $|t\rangle$ in terms of the quadrature states $|q\rangle$ as in (\ref{eq:gaussS}). The integral in $t$ cannot be exactly solved, but as we are interested in states with high levels of squeezing ($s$ tending to zero), the term in parenthesis can be approximated to $1+O(s^2)$, the terms of order 2 and higher being too small to be considered. The integral in $t$ is now
\begin{equation} \label{eq:GaussDJ8}
	\hat{\Pi}_{\Delta s}\left|\mathbf{S}\right\rangle\approx \pm\frac{1}{\pi}\int\mathrm{d}q\left(\int^{s+\Delta s/2}_{s-\Delta s/2} \frac{\mathrm{d}t}{t^2}e^{-q^2/t^2}\right)\left|q\right\rangle,
\end{equation}
which can be solved in terms of Gauss error functions \cite{erf72,erf08}, finally yielding
\begin{equation} \label{eq:GaussDJ9}
	\hat{\Pi}_{\Delta s}\left|\mathbf{S}\right\rangle=\pm\frac{1}{\sqrt{4\pi}}\int \mathrm{d}q\:\Gamma_{\text{err}}(q,s)\left|q\right\rangle,
\end{equation}
in the limit of high squeezing. The error function $\Gamma_{\text{err}}(q,s)$ is given by
\begin{equation} \label{eq:Ferr}
	\Gamma_{\text{err}}(q,s)=\frac{1}{q}\left\{\mathrm{erf}\left(\frac{q}{s-\Delta s/2}\right)-\mathrm{erf}\left(\frac{q}{s+\Delta s/2}\right)\right\}.
\end{equation}
These Gauss error functions arise naturally, measuring the quality of the Gaussian signal in terms of the quality of squeezing associated to $s$. If the function is constant indeed, we should measure the initial state with total certainty in the case of infinite squeezing as in (\ref{eq:CVDJ6}), but due to the realistic Gaussian states used in practice, the probability of measuring the initial state is given by $|\left\langle s\right|\!\left.\mathbf{S}\right\rangle|^2$, and then solving the resulting Gaussian integrals using equation (\ref{eq:GaussDJ4}) we obtain a probability of $\frac{1}{2}$. This means that the determinism of the algorithm is lost when using realistic states and the algorithm becomes probabilistic, when the function is constant we have a $50\%$ chance of getting the correct measurement at the output, we are no longer certain of the nature of the function with just one evaluation of the algorithm. However, its exponential speedup over a classical algorithm still holds in this case.

\section{Conclusions} \label{sec:concl}
Optical quantum computation has been proven a wide field of possibilities for practical implementations of quantum algorithms. We explored the combined action of discrete qubits and CV states of light to perform the well known Deutsch-Jozsa algorithm, taking advantage of both schemes of computation and using only optical devices. In previous implementations of the algorithm, the main problem lies in the difficulty of producing the oracle operation. The hybrid model proposed here alleviates the problem by using classical feed-forward to create an ancillary state that performs the desired operation.
		
As we have emphasized before, the importance of this particular algorithm lies in its relative simplicity that enables us to implement it in many settings. The hybrid proposal that we present also tries to inspire the implementation of algorithms based on the Fourier transform, with a nontrivial oracle, in order to find the possible new capabilities that such algorithms would acquire in contrast to the available implementations. For example, Shor's algorithm for integer factorization satisfies both conditions for the optical hybrid model.

In addition, we showed how the performance of the algorithm is diminished by considering (highly squeezed) Gaussian states in Alice's register. Hence, in a realistic implementation, apart from measurement errors, we would not have the ideal evolution of the quadrature states $|q\rangle$ either. Such implementation should provide some form of error correction together with repeated applications of the algorithm for a successful, fault-tolerant computation that does not depend highly on the quality of squeezing. We sense that a better measurement stage would help to solve the problem. 
		
On the other hand, other than all-optical implementations of hybrid computation exist, as the interaction of trapped atoms with a cavity mode of electromagnetic field \cite{Hybr08}. These proposals may be incorporated together in a ``super-hybrid'' computer, and take advantage of all the schemes to achieve scalability in the computation.

\nocite{*}
\bibliography{bibHDJ}
\end{document}